\documentclass[twocolumn,preprintnumbers,amsmath,amssymb]{revtex4}

\usepackage{graphicx}
\usepackage{dcolumn}
\usepackage{bm}

\begin{document}

\title{Anisotropic low-temperature piezoresistance in (311)A GaAs two-dimensional holes}

\author{B.~Habib, J.~Shabani, E.~P.~De~Poortere, M.~Shayegan}
\affiliation{Department of Electrical Engineering, Princeton
University, Princeton, NJ 08544, USA}

\author{R.~Winkler}
\affiliation {Department of Physics, Northern Illinois University,
De Kalb, IL, 60115, USA}

\date{\today}

\begin{abstract}
We report low-temperature resistance measurements in a
modulation-doped, (311)A GaAs two-dimensional hole system as a
function of applied in-plane strain. The data reveal a strong but
anisotropic piezoresistance whose magnitude depends on the density
as well as the direction along which the resistance is measured. At
a density of $1.6\times10^{11}$~cm$^{-2}$ and for a strain of about
$2\times10^{-4}$ applied along [01$\bar{1}$], e.g., the resistance
measured along this direction changes by nearly a factor of two
while the resistance change in the [$\bar{2}$33] direction is less
than 10\% and has the opposite sign. Our accurate energy band
calculations indicate a pronounced and anisotropic deformation of
the heavy-hole dispersion with strain, qualitatively consistent with
the experimental data. The extremely anisotropic magnitude of the
piezoresistance, however, lacks a quantitative explanation.
\end{abstract}

\maketitle

The piezoresistance effect -- the change of a device's electrical
resistance as a function of applied stress -- is of great
technological importance as it is utilized in making force,
displacement and pressure sensors \cite{Marian92}. It has also been
used recently in measuring sub-nanometer displacements of tips in
atomic force microscopes \cite{TortoneseAPL93, HarleyAPL99}, and
proposed data storage devices based on piezoresistive readback
\cite{ChuiAPL96}.

Strain affects the resistance of a solid in two ways. It changes the
physical dimensions of the resistor and can also modify its energy
band structure. In metals, it is the former effect that is dominant
and the the gauge factor, defined as the fractional change in
resistance divided by the solid's fractional change in length, is
dictated by the Poisson's ratio which is typically around 2. On the
other hand, for certain semiconductors the modification of the band
structure is a more prominent factor affecting the piezoresistance.
In multi-valley semiconductor systems, for example, strain changes
the relative population of the valleys. If the valleys have
anisotropic effective masses, the charge transfer induces a large
resistance change in certain crystallographic directions
\cite{Smith54}. Gauge factors of $\sim$1500 have been reported for
n-type Si inversion layers \cite{DordaSSC72} and $\sim$10,000 in
AlAs two-dimensional (2D) electron systems at low temperatures
\cite{ShkolnikovAPL04}.

In this Letter, we report a large, density (\emph{p}) dependent
piezoresistance effect in a (311)A GaAs 2D hole system (2DHS). We
utilize a simple but powerful technique to apply quantitatively
measurable in-plane strain \emph{in-situ} \cite{ShayeganAPL03} and
measure resistance parallel and perpendicular to the applied strain.
We deduce a gauge factor of $\sim$ 3600 at $p =
1.6\times10^{11}$~cm$^{-2}$ and $T = 0.3$~K. The data can be
qualitatively understood by the change in the heavy-hole valence
band shape with applied strain. But a quantitative explanation of
the observed piezoresistance magnitude and particularly its extreme
anisotropy, is lacking.

Our sample is grown on a (311)A GaAs substrate by molecular beam
epitaxy and contains a modulation-doped 2DHS confined to a
GaAs/AlGaAs heterostructure. The Al$_{0.35}$Ga$_{0.65}$As/GaAs
interface is separated from a 17~nm-thick Si-doped
Al$_{0.35}$Ga$_{0.65}$As layer (Si concentration of $4 \times
10^{18}$~cm$^{-3}$) by a 30~nm Al$_{0.35}$Ga$_{0.65}$As spacer
layer. We fabricated L-shaped Hall bar samples via photo-lithography
and used In:Zn alloyed at 440$^\circ$C for the ohmic contacts. Metal
gates were deposited on the sample's front (10nm Ti; 30nm Au) and
back (100nm Ti; 30nm Au) to control the 2D hole density.
Magneto-resistance data taken at $p = 2.1\times 10^{11}$~cm$^{-2}$
and $T = 0.3$~K yield mobilities of $1.7 \times 10^5$~cm$^2$/Vs and
$4.3 \times 10^5$~cm$^2$/Vs in the [$01\bar{1}$] and [$\bar{2}33$]
directions respectively. This mobility anisotropy in the (311)A
growth direction stems from the quasi-periodic ridges along the
$[\bar{2}33]$ direction \cite{WassermeierPRB95, HeremansJAL94}.

\begin{figure}
\centering
\includegraphics{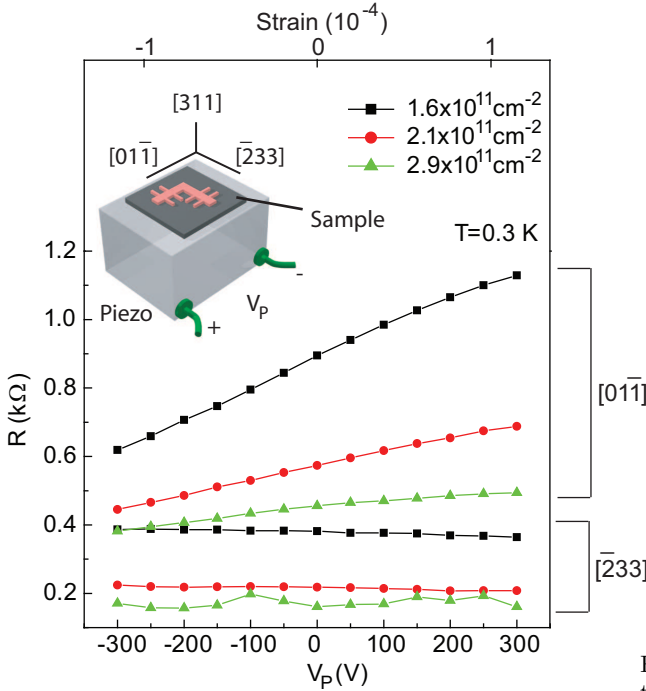}
\caption{\label{fig:M369L2pzr} (Color on-line) Piezoresistance
measured along the [01$\bar{1}$] and [$\bar{2}$33] directions for
three densities. To obtain the resistivity (in units of k$\Omega$ per square),
 multiply \emph{R} by 0.3. Inset: Experimental setup. The
poling direction for the piezo is along [$01\bar{1}$]. The strain
values indicated in the top scale have an uncertainty in the
absolute value and hence specify the change in applied strain
(see Ref. \cite{notecooldown}).}
\end{figure}

We apply tunable strain to the sample (thinned to $\sim200~\mu m$)
by gluing it on one side of a commercial multilayer piezoelectric
(piezo) actuator with the sample's [$01\bar{1}$] crystal direction
aligned with the poling direction of the piezo
(Fig.~\ref{fig:M369L2pzr} inset) \cite{ShayeganAPL03}. When bias
$V_P$ is applied to the piezo, it expands (shrinks) along the
[$01\bar{1}$] for $V_P > 0$ ($V_P < 0$) and shrinks (expands) along
the [$\bar{2}33$] direction. We have confirmed that this deformation
is fully transmitted to the sample, and using metal strain gauges
glued to the opposite side of the piezo, have measured its magnitude
\cite{ShayeganAPL03, GunawanPRL06}. Based on our calibrations of
similar piezo actuators, we estimate a strain of $3.8 \times
10^{-7}$~V$^{-1}$ along the poling direction. In the perpendicular
direction, the strain is approximately $-0.38$ times the strain in
the poling direction \cite{ShayeganAPL03,notepiezo}. In this paper
we specify strain values along the poling direction; we can achieve
a strain range of about $2.3 \times 10^{-4}$ by applying $-300~ \leq
V_P \leq 300$~V to the piezo \cite{notecooldown}. Finally, the
back-gate on the sample is kept at a constant voltage (0~V)
throughout the measurements to shield the 2DHS from the electric
field of the piezo actuator.

Figure \ref{fig:M369L2pzr} summarizes our measured piezoresistance.
Three trends in the data are clear. First, the change in resistance
(\emph{R}) with strain is density dependent and is larger for lower
densities. Second, the piezoresistance is more prominent along the
[01$\bar{1}$] direction. Indeed, at $p=1.6\times10^{11}$~cm$^{-2}$,
\emph{R} along this direction changes by a remarkable factor of two
for an applied strain of only $\sim2\times10^{-4}$. Third, in the
entire strain range we have studied, \emph{R} along [01$\bar{1}$] is
always larger than in the [$\bar{2}$33] direction.

\begin{figure}
\centering
\includegraphics{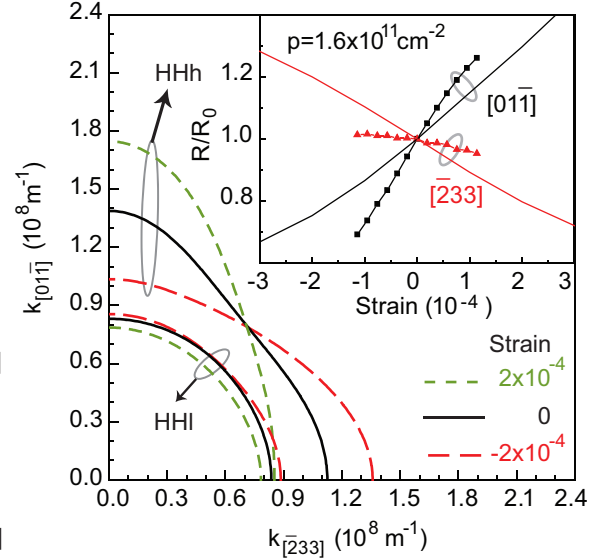}
\caption{\label{fig:M369L2con} (Color on-line) Calculated Fermi
contour plots for the HHl and HHh spin subbands; $k_{[01\bar{1}]}$
and $k_{[\bar{2}33]}$ are the wavevectors along the two
perpendicular ([01$\bar{1}$] and [$\bar{2}$33]) directions. Inset:
Comparison between the experimentally measured (symbols) and
calculated (lines) piezoresistance.}
\end{figure}

In order to understand the data of Fig.~\ref{fig:M369L2pzr}, we
performed self-consistent calculations of the Fermi contours of the
2D holes residing in the heavy-hole (HH) valence band as a function
of strain. Note that in this system the spin degeneracy of the HH
band is lifted due to the presence of strong spin-orbit coupling
\cite{Winkler03}. The spin subbands are termed heavy HH (HHh) and
light HH (HHl), reflecting the magnitude of their effective masses.
We used the $8\times8$ Kane Hamiltonian, augmented by the strain
Hamiltonian of Bir and Pikus \cite{Winkler03, Bir74}, to take into
account the spin-orbit coupling of the system as well as the
strain-induced contributions. We adapted this model to the (311)
orientation of our sample by a suitable coordinate transformation.
We note that these energy band calculations are quite accurate and
have quantitatively explained the spin-orbit induced spin-splitting
and its dependence on density, electric field and strain in 2DHSs
\cite{LuPRL98, PapadakisPHE01, HabibPRB07}.

The results of the energy band calculations for our sample
parameters are shown in Fig. \ref{fig:M369L2con}. Since all the
densities show similar trend, we only discuss results for
$p=1.6\times10^{11}$~cm$^{-2}$. Figure \ref{fig:M369L2con} shows
that the anisotropy of the HHh band is strongly enhanced with the
application of strain. This results in the anisotropy of the HHh
effective mass as well. Qualitatively, a larger Fermi wavevector in
a particular direction typically implies a larger mass in the same
direction. Hence, applying tensile strain along the [01$\bar{1}$]
direction leads to a larger mass along [01$\bar{1}$] and a smaller
mass along [$\bar{2}$33]. This change in effective mass
qualitatively explains the change in resistance in the two
directions.

To make a direct comparison with the experimental data, we
calculated the resistance for our sample parameters taking the
strain dependence of the anisotropy into account. We assume a
constant and isotropic relaxation time ($\tau$) and sum over spin
subbands, $n$, with energy dispersion $E_n (\mathbf{k})$. We
evaluate the velocity, $\partial E_n / (\hbar\,\partial k)$ at $E_F$
and $T=0$ to obtain the conductance ($\sigma$) \cite{Chambers90},

\begin{equation}
  \label{eqn:conduct}
  \sigma_{ij} = \tau e^2 \sum_n \int \!\frac{d^2k}{(2\pi)^2} \:
  \frac{\partial E_n}{\hbar\,\partial k_i} \:
  \frac{\partial E_n}{\hbar\,\partial k_j} \:
  \delta [E_F - E_n (\mathbf{k})].
\end{equation}
Since it is difficult to estimate the value of $\tau$, we normalize
the calculated resistance to its value at zero strain and compare
the calculations to the data of Fig. \ref{fig:M369L2pzr}. For the
comparison shown in the inset of Fig. \ref{fig:M369L2con}, the
experimental resistance was also normalized to the resistance at
zero applied piezo bias, $R_0$. As the figure shows, the
calculations certainly match the qualitative trend seen in the data.
It should be noted that the calculations are based on the particular
sample structure and density but contain no fitting parameters.

\begin{figure}
\centering
\includegraphics{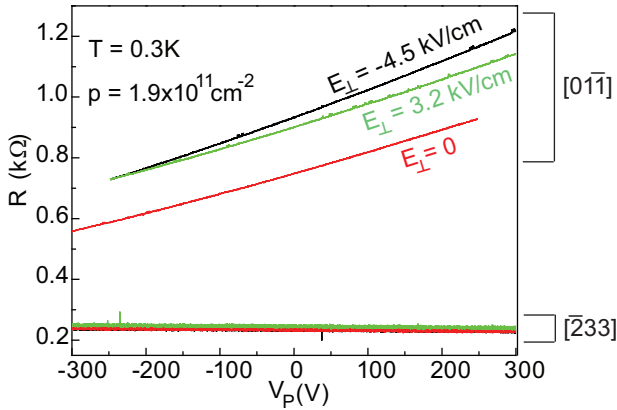}
\caption{\label{fig:M340H2pzef}(Color on-line) Electric field
dependence of piezoresistance in a 20~nm square well sample grown on
a (311)A surface at constant density. The gate biases are: (V$_F$ =
-0.2~V, V$_B$ = 0~V) for $E_\perp$ = -4.5~kV/cm, (V$_F$ = -0.375~V,
V$_B$ = 84~V) for $E_\perp$ = 0, and (V$_F$ = -0.5~V, V$_B$ = 144~V)
for $E_\perp$ = 3.2~kV/cm.}
\end{figure}

A puzzling aspect of Fig. \ref{fig:M369L2con} (inset) data is that
in the [01$\bar{1}$] direction, the experimental change in
resistance is typically larger than predicted by the calculations.
The [$\bar{2}$33] direction, on the other hand, exhibits a much
smaller change in the measured resistance than in the calculations.
It is tempting to relate this mismatch to how the quasi-periodic
corrugations \cite{WassermeierPRB95} parallel to [$\bar{2}$33] might
affect the directional dependence of $\tau$ with strain. However,
our piezoresistance measurements on a 20~nm square quantum well
sample, as a function of perpendicular electric field ($E_\perp$) at
a constant density, indicate otherwise [Fig. \ref{fig:M340H2pzef}].
The square well sample is glued to the piezo with the poling
direction in the [$01\bar{1}$] crystal direction. $E_\perp$ is tuned
while keeping the density constant with the help of front ($V_F$)
and back ($V_B$) gate biases \cite{noteQW}. At $E_\perp = 0$, the
confining potential of the square well and the carrier wavefunction
are symmetric. For $E_\perp \neq 0$, the potential becomes
asymmetric, resulting in the carriers being `pushed' closer to the
interface. Hence the resistance increases with the increase in the
magnitude of $E_\perp$ because of interface roughness, as shown in
Fig. \ref{fig:M340H2pzef} at a particular piezo bias. If the
direction dependent mismatch of the piezoresistance were related to
the corrugations along [$\bar{2}$33], we would expect a larger
piezoresistance mismatch with an increase in the magnitude of
$E_\perp$. On the contrary, Fig. \ref{fig:M340H2pzef} shows that the
piezoresistance is nearly independent of $E_\perp$ \cite{notecont}.

Moreover, a similar mismatch was also observed in (100) GaAs 2DHSs
\cite{KolokolovPRB99}, where the change in resistance was reported
to be larger than the calculated values in the [011] direction,
while along [01$\bar{1}$] the measured change was much smaller than
predicted by the calculations. The mobility anisotropy in (100) GaAs
2D holes at zero strain is only about 1.2-1.3 between the two
perpendicular ([01$\bar{1}$] and [011]) in-plane directions. This is
much smaller than the (311)A samples where the mobility anisotropy
is typically about 2 to 3 between the [01$\bar{1}$] and
[$\bar{2}$33] directions. It is unlikely, therefore, that the
corrugations in (311)A samples are responsible for the mismatch
between the rate of change of resistance along the [01$\bar{1}$] and
[$\bar{2}$33] directions. We emphasize that this mismatch is
independent of the poling direction of the applied strain: Our
measurements on a different heterostructure sample from the same
wafer as the one used in Fig. \ref{fig:M369L2pzr} but glued on a
piezo actuator with the poling direction along the [$\bar{2}$33]
direction also exhibit a larger change in resistance in the
[01$\bar{1}$] direction compared to [$\bar{2}$33].

In conclusion, we observe a large piezoresistance effect in (311)A
GaAs 2DHSs. The data indicate a strong dependence on density, and on
the in-plane direction along which the resistance is measured. The
maximum strain gauge factor we measure (along [01$\bar{1}$] and for
$p=1.6\times10^{11}$~cm$^{-2}$) is $\sim3600$. This value is
comparable to the ones reported, at liquid He temperatures, for
p-type Si (110) inversion layers at $p=5\times10^{11}$~cm$^{-2}$
\cite{DordaCON72} and for GaAs 2DHS grown in the (100) direction at
$p=7.6\times10^{11}$~cm$^{-2}$ \cite{KolokolovPRB99}.

We thank the ARO, DOE and NSF for support, and M. Grayson for
stimulating discussions.


\begin{thebibliography}{24}
\bibitem{notepiezo}
\bibinfo{note}{The strain in the perpendicular direction is taken into account
  in our calculations described later in the paper.}
\bibitem{notecooldown}
\bibinfo{note}{In our experiments, we can determine the relative changes in values of
  strain accurately, but not the piezo-bias corresponding to zero-strain
  precisely. This is because, thanks to a mismatch between the thermal
  expansion coefficients of GaAs and the piezo-stack, at low temperatures the
  sample can be under finite strain even at $V_P = 0$. This residual strain is
  cooldown-dependent and we do not know its precise value for our data. Based
  on our experience with cooldowns of samples glued to similar piezo-stacks, we
  expect a residual strain up to about $\pm1 \times 10^{-4}$.}
\bibitem{notecont}
\bibinfo{note}{This independence of the piezoresistance on $E_\perp$ is
  corroborated by our calculations of the Fermi contours as a function of
  $E_\perp$ for our quantum well.}
\end{thebibliography}
\end{document}